\documentclass[aps,twocolumn,nofootinbib]{revtex4}

\usepackage{amsmath}
\usepackage{amssymb}
\usepackage{bbm}
\usepackage{simplewick}
\usepackage{mathtools}
\usepackage[normalem]{ulem}
\usepackage{dsfont}
\usepackage{mathrsfs}
\usepackage[makeroom]{cancel}
\usepackage{graphicx}
\usepackage{bm}%ważne: aby tilde nad bold symbolami bylo OK

\def\0{\boldsymbol{0}}
\def\RR{\mathbbm{R}}

\def\xx{\boldsymbol{x}}
\def\kk{{\boldsymbol{k}}}
\def\rr{\boldsymbol{r}}
\def\yy{{\boldsymbol{y}}}
\def\zz{\boldsymbol{z}}

\newcommand{\Rzymskie}[1]{%
  \textup{\uppercase\expandafter{\romannumeral#1}}%
}

\def\A{\boldsymbol{A}}

\def\V{\boldsymbol{V}}
\def\E{\boldsymbol{E}}
\def\EE{\boldsymbol{{\cal E}}}

\def\Lag{{\cal L}}
\def\ii{\mathrm{i}}

\def\dd#1{d^3\mkern-1.5mu#1\,}

\def\bnabla{{\boldsymbol\nabla}}
\def\bpi{{\boldsymbol\pi}}

\def\B#1{\left(#1\right)}
\def\BB#1{\left[#1\right]}

\def\be{\begin{equation}}
\def\ee{\end{equation}}

\def\for{\ \text{for} \ }

\def\XXint#1#2#3{{\setbox0=\hbox{$#1{#2#3}{\int}$}
\vcenter{\hbox{$#2#3$}}\kern-.5\wd0}}

\usepackage[ddmmyyyy]{datetime}
\makeatletter
\def\Dated@name{}
\makeatother

\def\GE{G_{\cal E}}
\def\bAE{\A_{\cal E}}
\def\AE{A_{\cal E}}

\def\PhiE{\Phi_{\cal E}}
\def\fE{f_{\cal E}}
\def\gE{g_{\cal E}}
\def\FE{\boldsymbol{F}_{\cal E}}

\def\DiV{{\bnabla\cdot}}

\begin{document}
\title{Electric field-based quantization of the gauge invariant Proca theory}
\author{Bogdan Damski}
\affiliation{Jagiellonian University,
Faculty of Physics, Astronomy and Applied Computer Science,
{\L}ojasiewicza 11, 30-348 Krak\'ow, Poland}
\begin{abstract} 
We consider the gauge invariant version of the Proca theory, 
where besides the real vector field there is also the 
real scalar 
field. 
We quantize the theory such that  the commutator
of the scalar field operator and the electric field operator
is given by a predefined three-dimensional 
vector field, 
say $\EE$ up to a
global prefactor. This happens when 
the field operators of the gauge invariant 
Proca theory satisfy the
proper gauge 
constraint.
In particular, we show that $\EE$ 
given by the  classical Coulomb field leads to 
the Coulomb gauge constraint making 
the vector field operator divergenceless.
We also show that physically unreadable gauge constraints 
can  have a strikingly simple  $\EE$-representation 
in our formalism. This leads to the 
discussion of Debye, Yukawa,  etc. 
gauges.
In general terms, we explore  the mapping 
between  classical vector 
fields and  gauge  constraints  
imposed on the operators of 
the studied theory.
\end{abstract}
%\date{\today}
\maketitle

\section{Introduction}
\label{Introduction_sec}

The Proca theory delivers  the simplest 
relativistic description of
massive vector bosons 
\cite{Greiner,Nieto_RMP2010}. 
As a result of that, 
it is of both phenomenological 
and theoretical interest.

In the phenomenological context, 
it captures  some  properties of 
$\rho$ and $\omega$
mesons and 
the particles 
mediating  weak interactions, 
W and
Z bosons \cite{Greiner}.
In addition to that, it is  
regarded as a promising 
extension of
 Maxwell's electrodynamics, 
 the one taking into 
 account the possibility that 
 the photon may  not  
be a  massless particle after all. 
Thereby various upper bounds on the 
photon mass are obtained by comparing 
the predictions of the Proca theory 
 to actual experimental data 
(see e.g.  \cite{Gillies2005,Nieto_RMP2010}
extensively discussing
this  physically rich  topic).  
In the   theoretical context,
which is of main interest in this work,
the Proca theory  
provides an elegant framework 
for the  examination of various 
issues associated with the 
quantization of vector fields 
(see e.g. \cite{Greiner,ColemanBook,Weinberg}).

We are interested in the  Proca theory of the real 
vector field. Its classical  Lagrangian density
can be written as 
\be
\Lag=-\frac{1}{4} 
\B{\partial_\mu V_\nu-\partial_\nu V_\mu}^2
+\frac{m^2}{2}\B{V_\mu}^2,
\label{SPr}
\ee
where $V^\mu$ is the vector field 
and $m$ is the mass of spin-1
particles described by this theory after its 
quantization 
(see Appendix  for our conventions).

The important thing now is that theory
(\ref{SPr}) is manifestly non-invariant 
with respect to the 
 gauge transformation. 
In fact, it   is a gauge-fixed theory in the sense that 
 field equations impose the Lorenz gauge
constraint onto the vector field.
This state of affairs can be easily changed by
the replacement 
\be
V_\mu\to A_\mu + \frac{1}{e}\partial_\mu G,
\label{stuck}
\ee
where  the  vector field  
$A^\mu$ and the real scalar field
$G$  are supposed to simultaneously 
change under the 
gauge transformation. Namely, 
\be
A_\mu\to A_\mu +\partial_\mu f,  \ G \to G  - e f,
\label{gauge12}
\ee
where $f$ is a smooth
real  function  of space-time coordinates and 
$e$ is the unit of the  electric charge.

Imposing (\ref{stuck}) on (\ref{SPr}), we see  that 
the resulting Lagrangian density,
\be
\Lag'=-\frac{1}{4} \B{\partial_\mu A_\nu-\partial_\nu A_\mu}^2
+\frac{m^2}{2}\B{A_\mu + \frac{1}{e}\partial_\mu G}^2,
\label{S2}%
\ee
is unaffected by 
the  gauge transformation.
For this reason, we will refer to the theory 
defined by 
(\ref{S2}) as the gauge invariant (GI) Proca theory. 
Such a theory was  studied before,
see e.g. \cite{Guralnik1968,Pimentel2015},
and 
it bears  similarity to the 
Stueckelberg theory, which is 
reviewed in \cite{Ruegg_IJMPA2004}.

To proceed with  the discussion of the GI Proca theory, 
one has to choose a gauge because the vector field 
is
no longer Lorenz gauge fixed in (\ref{S2}). 
Besides the standard Coulomb gauge choice, 
which was e.g. enforced
 with the Lagrange
 multiplier technique in \cite{Guralnik1968},
 the following
intriguing
gauge constraint was introduced in \cite{Pimentel2015}
\be
e\DiV\A_D=m^2 G_D.
\label{Yuk_cons}
\ee
It was labelled  as the Coulomb gauge
choice \cite{Pimentel2015},
but the rationale behind such a name was not provided.
We believe that a proper name for such a gauge 
could be  the 
Debye  gauge, which will be carefully explained in 
this work.
Anticipating this discussion,
we have  labelled the fields subjected to such a 
constraint with the appropriate subscript.
Their  quantization was  studied  in
\cite{Pimentel2015} by means of the 
 Faddeev-Jackiw
 approach  \cite{FaddeevJackiwPRL1988}.

Our goal  is to develop and discuss 
the quantization formalism, where  gauge choices
are labelled by the classical vector 
field $\EE$, 
which determines  the commutator of the scalar
and electric 
field operators.
Thereby, we    explore the 
 mapping between  such $\EE$
and the 
field operators of the GI Proca theory.

The outline of this paper is the following.
The concise summary of 
basic results
concerning the  Proca theory is
provided in Sec. \ref{Basics_sec}.
Next, our  quantization
procedure  is  introduced
in Sec. \ref{Gauge_sec}.
Its features  are then    
 discussed  in Sec. \ref{Electric_sec},
where the  electric field context of the 
proposed approach is 
laid out along with
several 
illustrative   examples.
Finally, the  summary of our work
is  presented 
in Sec. \ref{Summary_sec}, which is followed by 
Appendix   listing our conventions.

%\clearpage

\section{Basics}
\label{Basics_sec}
We state below basic  results
concerning  theories (\ref{SPr}) and (\ref{S2}).

To begin, the independent
variables of
Proca theory (\ref{SPr})
are fields $V^i$ and their 
canonical conjugates 
\be
\pi_i=\partial_i V_0 -\partial_0 V_i.
\label{sssq1}
\ee
Such  a theory  is canonically
quantized by demanding that  \cite{Greiner,ColemanBook,Weinberg}
\begin{align}
\label{CaNon}
&[V^i(t,\xx),\pi^j(t,\yy)]=-\ii\delta^{ij}\delta(\yy-\xx),\\
&[V^i(t,\xx),V^j(t,\yy)]=[\pi^i(t,\xx),\pi^j(t,\yy)]=0.
\label{CaNon1}
\end{align}
We note that 
\be
V^0=-\frac{1}{m^2}\DiV\bpi,
\label{V000}
\ee
which explains why $V^0$ is the dependent variable 
of theory (\ref{SPr}). We also 
note that the canonical conjugate of 
$V^0$ vanishes.

Then, we remark that  the 
 variables of  GI Proca
theory (\ref{S2}), whose quantization will be
discussed in Sec. \ref{Gauge_sec}, 
are fields $A^i$ and $G$
as well as their canonical conjugates   
\be
\partial_i A_0 -\partial_0 A_i=\pi_i
\label{sssq11}
\ee
and 
\be
\tilde{\pi}=\frac{m^2}{e}\B{A_0+\frac{1}{e}\partial_0 G}=\frac{m^2}{e}V_0,
\label{sssq2}
\ee
respectively. We note that 
the right-hand sides of 
(\ref{sssq11}) and (\ref{sssq2}) 
follow from 
mapping (\ref{stuck}), which we 
assume in this work.

Finally, we have a few observations 
about $\bpi$ and $\tilde{\pi}$. 
First, (\ref{V000}) and 
(\ref{sssq2})  imply that 
$\bpi$ and    $\tilde{\pi}$
 are  linked via the 
{\it field constraint} 
\cite{RemarkFieldConstraint}
\be
\DiV\bpi=-e\tilde{\pi}.
\label{field_cons}
\ee 
Second,   $\bpi$ and    $\tilde{\pi}$ 
are gauge invariant.
This means that unlike $A$ and $G$,
they  will not be 
equipped 
with  a gauge-specific subscript below.
Third, the
physical content of 
$\bpi$, and so also of    $\tilde{\pi}$
due to (\ref{field_cons}), 
  is best seen from the fact that  
$\bpi=\E$, where 
$\E=-\partial_0\V-\bnabla V^0=
-\partial_0\A-\bnabla A^0$ 
is 
the electric field operator.
Note that we  use   the same 
``electric field'' terminology as
in the theory of the massless 
electromagnetic  field.

%\clearpage

\section{Gauge ansatz and commutation relations}
\label{Gauge_sec}

We are interested  in 
quantization of theory (\ref{S2}).
In a nutshell, 
one may approach 
 this problem 
 in the  following 
 way.

To begin, one chooses 
the gauge constraint for
the fields.
For example, one may decide to work
in the 
Coulomb gauge
\be
\DiV\A_C=0,
\label{Cou_cons}
\ee
where the subscript indicates the gauge choice.
Naturally, there are uncountably 
many  other  
gauge choices, see e.g.  (\ref{Yuk_cons}), 
whose implications are  not so 
obvious.

Then, one figures out
commutation relations between the 
fields and their canonical  conjugates,
which is a non-trivial task. Indeed, 
as they  
have to be  consistent
with the chosen  gauge constraint,
they are expected to 
differ from the canonical 
commutation relations.

We approach quantization 
of theory (\ref{S2})
somewhat differently. Namely, 
instead of imposing the 
specific gauge constraint in the
form of the   equation 
for the vector  and 
scalar field operators, we 
require
that  
\begin{subequations}
\begin{align}
\label{chigauge}
&\GE(t,\xx) = e\int \dd{z} \V(t,\zz)\cdot \EE(\zz-\xx),\\
&\bAE=\V+\frac{1}{ e}\bnabla \GE,
\label{Avarphi}
\end{align}
\label{gauge_ansatz}%
\end{subequations}
where $\EE$ is a time-independent 
$\RR^3$-valued
vector field
and the  appropriate subscript has been added to the
fields to indicate their dependence on $\EE$.
Equation (\ref{chigauge}) can be seen as the ansatz, 
whereas equation (\ref{Avarphi}) 
expresses the fact that we rely on 
mapping (\ref{stuck}), which also 
leads to 
\be
\AE^0=
\frac{e}{m^2}\tilde{\pi}
-\frac{1}{e}\partial_0\GE.
\label{Ae00}
\ee
All together, we will
refer to  
(\ref{gauge_ansatz}) as  the {\it gauge ansatz}.

The field $\EE$, whose  meaning will be 
discussed in Sec. \ref{Electric_sec},
defines the gauge in our formalism.
In fact, it is easy to see that 
under $\EE\to\EE'$, $\GE$ and $\AE$ 
transform  just as $G$ and $A$
in (\ref{gauge12}) with
\be
f(t,\xx)=\int\dd{z}\V(t,\zz)\cdot
[\EE(\zz-\xx)-\EE'(\zz-\xx)].
\ee
This time, however, $f$ is operator-valued.
This is interesting because classical, i.e. $c$-number,  
gauge transformations are typically discussed 
in the context of 
gauge theories (see e.g. Sec. 2.5.2 of \cite{Leader2014}
for relevant remarks).

We are now ready  to discuss
equal-time commutators between 
the canonically-related operators
introduced in Sec. \ref{Basics_sec}.
The non-trivial  ones
are 
\begin{align}
\label{G1}
&[\GE(t,\xx),\tilde{\pi}(t,\yy)]=\ii \,  \DiV\EE(\yy-\xx),\\
\label{G12}
&[\AE^i(t,\xx),\tilde{\pi}(t,\yy)]= \frac{\ii}{e}\partial^y_i[\delta(\yy-\xx)-\DiV\EE(\yy-\xx)],\\
\label{G3}
&[\GE(t,\xx),\pi^j(t,\yy)]= -\ii   e{\cal E}^j(\yy-\xx),\\
\label{G2}
&[\AE^i(t,\xx),\pi^j(t,\yy)]= -\ii\delta^{ij} \delta(\yy-\xx) + \ii
\partial_i^y {\cal E}^j(\yy-\xx),
\end{align}
where $\partial^y_i=\partial/\partial y^i$.
These expressions trigger the following comments.

First,  in order to verify these
 commutators, one can 
replace $\GE$ and $\AE^i$
in (\ref{G1})--(\ref{G2})
with (\ref{gauge_ansatz})
and then 
use  (\ref{CaNon})
 to simplify  the resulting 
expressions.
Similarly, one may verify with the 
help of (\ref{CaNon1}) that the remaining
equal-time 
commutators between
$\GE$, $\bAE$, $\bpi$, and $\tilde{\pi}$
  identically  vanish.

Second, we find these
commutators remarkably compact and general.
As expected, they 
do differ from canonical commutation relations:
(\ref{G1}) is not equal to $\ii\delta(\yy-\xx)$,
(\ref{G2}) is not equal to $-\ii\delta^{ij} \delta(\yy-\xx)$, and 
(\ref{G12}) as well as (\ref{G3}) do not vanish.
The  structure of (\ref{G1})--(\ref{G2}) stems 
from the restrictions imposed 
by field constraint (\ref{field_cons})
and 
gauge ansatz  (\ref{gauge_ansatz});
see 
Sec. \ref{G_sec} for additional relevant remarks.
In particular, one may 
easily notice that 
(\ref{G1}) and (\ref{G3}) 
are interrelated via (\ref{field_cons}).
The same remark applies to 
(\ref{G12}) and (\ref{G2}).

Third, we have independently verified 
the above results in the two already 
introduced gauges, 
 (\ref{Yuk_cons}) and
(\ref{Cou_cons}),
where $\EE$ is given by 
(\ref{Debye}) evaluated for 
$M=m$ and (\ref{Coulomb}),
respectively.
We have  done it
via the Dirac bracket quantization 
technique adopted so as  to
enforce  gauge  constraints
(\ref{Yuk_cons}) and
(\ref{Cou_cons}) 
(see \cite{Weinberg}
for the  textbook  introduction to such a 
quantization approach and 
Sec. \ref{Gradient_sub} for the explanation 
of the above-listed choices of $\EE$).

%\clearpage

\section{Electric field perspective on gauge ansatz}
\label{Electric_sec}
The quantum GI Proca theory is built of  
the vector field 
$\AE$ and the scalar field $\GE$. The role 
of $\AE$  is clear: the electric and magnetic 
field operators are expressed in terms of $\AE$, 
and so in such a sense this operator 
captures physics of  the electromagnetic field.
The question now is what is the role of 
 $\GE$?
At first sight, it seems that 
 the  only 
role of $\GE$ is to enforce the gauge invariance 
of the Lagrangian density.
However, by looking at commutator 
(\ref{G3}), we realize
that $\GE$ also plays 
the role of the generator
of the local  shift  
of the electric field operator.
To explain what we mean by saying
so, 
we note that by 
combining
(\ref{G3}) with the following well-known
identity
\begin{multline}
\exp(X)Y\exp(-X)  \\
= Y + [X,Y] + \frac{1}{2!}[X,[X,Y]] + \cdots,
\label{XYX}
\end{multline}
it can be formally  shown that 
\begin{multline}
\exp\BB{\ii\GE(t,\xx)}
\E(t,\yy)
\exp\BB{-\ii \GE(t,\xx)}\\
= \E(t,\yy) + e\EE(\yy-\xx).
\label{eeUpsilon}
\end{multline}
As both (\ref{G3}) and (\ref{eeUpsilon})
particularly clearly expose the electric field context of 
$\EE$, we see 
the quantization procedure 
based on  (\ref{gauge_ansatz})
 as the electric field-based quantization  
scheme. Two remarks are in order now.

First, we use the term formal when we refer 
to (\ref{eeUpsilon}) because we do not actually 
inquire  if
the operator $\exp[\pm\ii\GE(t,\xx)]$
is well-defined. 
Second,  we note  that in the 
spirit of the Helmholtz theorem \cite{GriffithsBook}, 
one may consider the following 
decomposition of $\EE$
\be
\EE=-\bnabla \PhiE + \bnabla\times \FE,
\label{OPOOPO}
\ee
where $\PhiE$ and $\FE$
are
classical time-independent
scalar and vector fields, 
respectively.
Formula (\ref{OPOOPO}) 
will guide our subsequent
discussion.

\subsection{Curl-free  $\EE$}
\label{Gradient_sub}
We  study here gauges induced by 
\be
\EE=-\bnabla\Phi_{\cal E},
\label{EErr}
\ee
where $\PhiE$ is real-valued.

To begin, we  address the
question of  what is  the relation between 
$\GE$ and $\bAE$ when  (\ref{EErr}) holds. 
After standard manipulations based on 
gauge ansatz (\ref{gauge_ansatz}), we find that 
\be
e\DiV\bAE=\fE(-\ii \bnabla)\GE+\Delta\GE,
\label{chjcdj}%
\ee
where $\fE$ is defined via
\be
\PhiE(\rr)=\int\frac{\dd{k}}{(2\pi)^3}\frac{\exp(-\ii\kk\cdot\rr)}{\fE(\kk)}
\label{PhifEee}
\ee
and  $\fE(\kk)=\fE^*(-\kk)$   
to ensure the real value of the above
integral.
We will refer to (\ref{chjcdj}) as the 
{\it gauge  constraint} to distinguish it from 
  field constraint (\ref{field_cons})
  and   gauge  ansatz (\ref{gauge_ansatz}). The 
  {\it formal} character of (\ref{chjcdj}) will
be commented upon  in Sec. \ref{Summary_sec}.
We are now ready to discuss 
the previously mentioned  Coulomb and Debye  gauges.

We say that $\EE$ induces the Coulomb gauge when 
\be
\EE=-\bnabla\Phi_C, \ \Phi_C=\frac{1}{4\pi r},
\label{Coulomb}
\ee
where 
$\bnabla=(\partial/\partial r^i)$
and  $r=|\rr|$.
Such a terminology is supported by two observations.
First, it is natural in our formalism because 
such $\EE$ is given by the negative gradient 
of the Coulomb potential originating from the unit 
charge.
Second, a simple calculation shows that 
$\fE(-\ii \bnabla)=-\Delta$ here, which 
 leads to the 
divergenceless vector 
field  via (\ref{chjcdj}).
Properly labeling the fields, we have   
\be
(G_C,\A_C)=(\GE,\bAE) \for \EE=-\bnabla\Phi_C,
\ee
where the vector field  satisfies
 gauge constraint (\ref{Cou_cons})  in the traditional
nomenclature.

In full analogy to the above   reasoning,
the gauge 
induced by 
\be
\EE=-\bnabla\Phi_D, \  
\Phi_D=\frac{\exp(-Mr)}{4\pi r}
\label{Debye}
\ee
will be called the  Debye  gauge ($M>0$).
We have  proposed this  name because such 
$\EE$ is given by the 
negative gradient of the Debye potential
describing the screening 
of the  unit 
charge in plasmas and electrolytes.

As far as the relation between $\GE$ and $\bAE$ is concerned,
we find
$\fE(-\ii \bnabla)=-\Delta+M^2$
 in the Debye gauge. 
Then, it follows from (\ref{chjcdj})
that the fields in such a  gauge satisfy
\be
e\DiV\A_{D} = M^2 G_{D}.
\label{D_cons}
\ee
Note that previously stated 
 gauge constraint (\ref{Yuk_cons}) is the 
$M=m$ version of (\ref{D_cons}).

Next, we observe that the
gauge constraint satisfied by the 
fields 
non-trivially depends on 
the
magnitude and the direction of
$\EE$  (the
magnitude and the sign of $\PhiE$). 
This can be illustrated by the introduction 
of the following two gauges.

We define the primed Coulomb 
gauge by saying that it is induced in our
formalism by 
\be
\EE=-\bnabla\Phi_{C'}, \ \Phi_{C'}=\beta\Phi_C=\frac{\beta}{4\pi r},
\label{CoulombP}
\ee
where  $\beta>0$.
The sensitivity of the 
 gauge constraint to the change of the magnitude 
 of $\EE$ is now seen by 
comparing (\ref{Cou_cons})
to 
\be
e\DiV\A_{C'}=\frac{\beta-1}{\beta}\Delta G_{C'},
\label{Cpr_cons}
\ee
which is satisfied by the fields in the primed 
Coulomb gauge.

Furthermore, 
we consider the gauge induced by 
\be
\EE=-\bnabla\Phi_Y, \
\Phi_Y=-\Phi_D=-\frac{\exp(-Mr)}{4\pi r},
\label{Yukawa}
\ee
where the subscript refers to the fact that 
such $\EE$ is given by the negative 
gradient of the Yukawa potential obtained 
for the
unit  strength of  the inter-nucleon 
interactions.
The  fields in so defined  Yukawa gauge 
satisfy  
\be
e\DiV\A_Y=2\Delta G_Y - M^2G_Y.
\label{Y_cons}
\ee
The difference between  (\ref{D_cons})
and (\ref{Y_cons}) 
illustrates the sensitivity  of
the gauge constraint to 
the global change of the direction of 
$\EE$.

Moving on, we  note that 
new gauges can be obtained by 
superposing 
fields  $\EE$.
For fields $\EE$ given by (\ref{EErr}),
this typically 
leads to the complicated
relation
 between
$\GE$ and $\bAE$ due to
 the reciprocal 
additivity law for  $\fE$. 
Namely, if
\be
\EE=-\bnabla\Phi_{{\cal E}'}-\bnabla\Phi_{{\cal E}''}-\cdots,
\label{EEsup}
\ee
then 
\be
\frac{1}{\fE}=\frac{1}{f_{{\cal E}'}}+\frac{1}{f_{{\cal E}''}}+\cdots.
\label{fErecip}
\ee

This can be illustrated by the 
 consideration of  the 
Coulomb-Yukawa gauge, which we define 
as the gauge induced by 
\be
\EE=-\bnabla\Phi_C-\bnabla\Phi_Y
=-\bnabla\B{
\frac{1}{4\pi r}
-\frac{\exp(-Mr)}{4\pi r}
}.
\label{EECY}
\ee
A quick calculation  shows that 
in this case 
$\fE(-\ii \bnabla)=(\Delta/M)^2-\Delta$,
which results  in 
\be
e \DiV\A_{CY}=\frac{1}{ M^2}
\Delta(\Delta G_{CY}).
\label{CY_cons}
\ee
Note that such a gauge constraint 
resembles neither (\ref{Cou_cons}) 
nor (\ref{Y_cons}) despite the fact 
that it is induced by the
superposition of 
the fields $\EE$ leading to 
the Coulomb and 
Yukawa gauges. This is 
the  consequence of 
the fact that 
$\fE\neq f_{{\cal E}'}+f_{{\cal E}''}+\cdots$
when  (\ref{EEsup}) holds.
We mention in passing 
that the $M=m$ version of the 
operator $G_{CY}$ 
 was used in \cite{BDPeriodic1}
to construct the finite-energy 
charged state in  Proca theory (\ref{SPr}).

Finally, we note  the trivial possibility of choosing
$\EE=\0$. This sets  $\GE=0$,  removing
the scalar field from the theory. 
Such a gauge choice is known in the literature as 
the unitary gauge (see e.g.  \cite{DasBookStueck}). In 
our formalism, the term null gauge  seems to be
more appropriate.

\subsection{Divergence-free  $\EE$}
\label{Curl_sub}
We briefly comment   here
upon  gauges induced by   
\be
\EE=\bnabla\times\FE, 
\label{curlll}
\ee
where $\FE$ is $\RR^3$-valued.

For a general 
function $\FE$, we are unsure  
how to  derive the closed-form 
expression for the gauge constraint
akin to (\ref{chjcdj}). 
Thus, we focus  on the 
specific results inspired by the 
discussion from Sec. \ref{Gradient_sub}.
Namely, we   consider 
\be
\FE(\rr)=\boldsymbol{d}
\int\frac{\dd{k}}{(2\pi)^3}\frac{\exp(-\ii\kk\cdot\rr)}{\gE(\kk)},
\label{FE11}
\ee
where  $\boldsymbol{d}\in\RR^3$ is 
the  constant vector and
$\gE(\kk)=\gE^*(-\kk)$.
It can be then  found via (\ref{gauge_ansatz}) 
that  the fields of the GI Proca theory 
satisfy the following  formal
gauge constraint
\be
e \boldsymbol{d}\cdot(\bnabla\times\bAE)=\gE(-\ii\bnabla)\GE.
\label{curl_fE}
\ee

To see how all this works in practice, 
one may   choose $\FE$ 
to be given by 
\be
\boldsymbol{d}\frac{\beta}{4\pi r}, \ 
\pm \boldsymbol{d} \frac{\exp(-Mr)}{4\pi r},
\ \boldsymbol{d}\B{\frac{1}{4\pi r}-\frac{\exp(-Mr)}{4\pi r}}.
\label{Curl_spec}
\ee
From the results presented in 
Sec. \ref{Gradient_sub}, 
it is clear  that
these choices lead to 
$\gE(-\ii\bnabla)$ equal to 
\be
-\frac{1}{\beta}\Delta, \ \pm(-\Delta+ M^2), \ 
(\Delta/M)^2-\Delta,
\label{ggEEgg}
\ee
respectively.
The corresponding gauge
constraints are 
obtained by 
combining (\ref{curl_fE}) 
with 
(\ref{ggEEgg}),
the $\EE$ fields associated with 
them  are given by the curl of 
the 
vector fields listed in (\ref{Curl_spec}).

\subsection{Gauge constraints
vs. commutation relations} 
\label{G_sec}

Let us consider  a gauge constraint
 written in the form $\Upsilon=0$.
We will say that it  is consistent with 
equal-time 
commutation relations, written for the fields 
belonging to some set ${\cal X}$, 
when $[\Upsilon(t,\xx), X(t,\yy)]=0$ 
for all $X\in{\cal X}$.
For example, the consistency of  
gauge constraint 
(\ref{D_cons}) with  
commutation relations
 (\ref{G1})--(\ref{G2}) requires  
 $[e\DiV\A_{D}(t,\xx) - M^2 G_{D}(t,\xx),X(t,\yy)]=0$
 for  $X=G_D, \A_D, \bpi, \tilde{\pi}$.

We note that it can be easily  verified  that 
gauge constraints (\ref{Cou_cons}), (\ref{D_cons}),
 (\ref{Cpr_cons}), (\ref{Y_cons}), and (\ref{CY_cons}) 
 are consistent with commutation relations 
 (\ref{G1})--(\ref{G2}). 
It goes without saying that this happens 
when the right-hand sides of
(\ref{G1})--(\ref{G2}) are evaluated with 
 the corresponding 
 fields $\EE$: (\ref{Coulomb}), (\ref{Debye}),  
 (\ref{CoulombP}), (\ref{Yukawa}), and (\ref{EECY}),
 respectively.
For a general
curl-free   $\EE$ given by (\ref{EErr}),
it can be formally shown that gauge constraint 
(\ref{chjcdj}) is consistent with (\ref{G1})--(\ref{G2}).

We also note that similar self-consistency 
checks can be performed for the
divergence-free $\EE$ discussed in Sec. \ref{Curl_sub}. 
Namely, it can be shown that 
(\ref{curl_fE}) is formally consistent with 
(\ref{G1})--(\ref{G2}) when (\ref{FE11}) holds,
which can be also individually 
verified for  the  specific cases listed in 
(\ref{Curl_spec}).

%\clearpage

\section{Summary}
\label{Summary_sec}

We have discussed how  GI  Proca theory (\ref{S2})
can be  quantized with the help of  gauge ansatz 
(\ref{gauge_ansatz}). Such an ansatz is parameterized 
by the classical vector field $\EE$, which 
determines the commutator of the scalar field 
operator  and  the electric field operator 
(\ref{G3}).

In several special cases, we have found
an explicit mapping between 
the field $\EE$ 
and the 
gauge constraint satisfied by
the  fields 
 of the GI Proca theory.
In particular, we have  discussed the mapping
\be
\EE=-\bnabla\B{\frac{1}{4\pi r}} \  \mapsto  \ \DiV \A_C=0,
\ee
which gives a new  meaning to the term Coulomb gauge,
  a very  suggestive one in our opinion.
While discussing other cases, we have 
found that unreadable gauge constraints
can have a strikingly   simple $\EE$-representation
 in our formalism, which we find remarkable.
One of the simplest illustrations
supporting such an observation is the following 
\be
\EE=-\bnabla\
\B{\frac{\exp(-Mr)}{4\pi r}} \  \mapsto \   e\DiV\A_D = M^2 G_D,
\ee
which  defines  the 
Debye gauge  in  our nomenclature.
Further support  for the above 
observation is provided 
 by
 comparing gauge constraints 
(\ref{Cpr_cons}), (\ref{Y_cons}),
and (\ref{CY_cons})  to the 
fields $\EE$ associated with  them 
(\ref{CoulombP}), (\ref{Yukawa}), and (\ref{EECY}),
 respectively.
We note that, to the best of our 
knowledge, none of these three 
gauge constraints has  been 
previously mentioned  in the 
literature.
We also note that another batch 
of unusual     gauge constraints,
having 
simple $\EE$-representation,
can be obtained by combining 
(\ref{curl_fE})
with (\ref{ggEEgg}).

In a more general context, we have proposed 
 the relation  between curl-free  $\EE$ given by 
(\ref{EErr}) and 
the gauge constraint satisfied by the fields of
the GI Proca theory (\ref{chjcdj}).
Such a result 
 has a formal character
because it involves the pseudo-differential 
operator $\fE(-\ii\bnabla)$, where
the function 
$\fE$ can be in general non-analytic or
singular for well-defined  $\EE$.
If such complications are present, then there 
is   the  question of what
 (\ref{chjcdj}) really means.
These somewhat intriguing ambiguities  
do not affect our 
gauge ansatz-based considerations (\ref{gauge_ansatz}),
which do not rely on the form of the gauge 
constraint satisfied by the fields.
Similar remarks apply to formal result
(\ref{curl_fE}),
which has been obtained for 
the
particular class of 
divergence-free    $\EE$.

Finally, we would like to emphasize the 
efficiency of the 
discussed formalism.
Indeed, our  quantization procedure is
carried out all at once for 
different gauges labelled by
$\EE$.
This is  illustrated 
by the general character of
commutation relations (\ref{G1})--(\ref{G2}).

\section*{ACKNOWLEDGMENTS}
These studies have
been  supported by the Polish National
Science Centre (NCN) Grant No. 2019/35/B/ST2/00034.
The research for this publication has been also supported 
by a grant from the Priority Research Area DigiWorld under
the Strategic Programme Excellence Initiative at Jagiellonian University.

\appendix*
\section{CONVENTIONS}
\label{App_Conv}

We adopt  the Heaviside-Lorentz system of units
in its $\hbar=c=1$ version.
Greek and Latin indices of tensors  take values $0,1,2,3$ and   $1,2,3$,
respectively. 
The metric signature is $(+---)$.
$3$-vectors are written in bold, e.g. $x=(x^\mu)=(x^0,\xx)$.
We use the Einstein summation convention,  
$(X_{\mu\cdots})^2=X_{\mu\cdots}X^{\mu\cdots}$,
    %$\Box=\partial\cdot\partial=\partial_\mu\partial^\mu$,
and $\Delta=\bnabla\cdot\bnabla$. 
The complex conjugation is denoted as $*$.

%\bibliography{../Scommon/reference.bib,paper_specific_references.bib} 

\begin{thebibliography}{14}
\bibitem[{Rem()}]{RemarkFieldConstraint}
\bibinfo{note}{We call (\ref{field_cons}) the field constraint because it
  follows from the field equations of the Proca theory.}
\end{thebibliography}

\end{document}